\documentclass[journal=ancac3,manuscript=article]{achemso}
\usepackage{chemformula} 
\usepackage{xr}
\usepackage[T1]{fontenc} 
\usepackage{graphicx}
\usepackage{adjustbox}
\usepackage{upgreek}
\usepackage[normalem]{ulem}
\usepackage{amsmath}
\usepackage{amssymb}
\usepackage{multirow}
\usepackage{multicol}

\usepackage{caption}

\SectionNumbersOn

\author{Ramesh Kudalippalliyalil}
\altaffiliation{RK and GP contributed equally to this work}
\affiliation[UMD]
{Institute for Research in Electronics and Applied Physics; Department of Electrical and Computer Engineering, University of Maryland, College Park, MD}

\author{Gyan Prakash}
\altaffiliation{RK and GP contributed equally to this work}
\affiliation[UMD]
{Institute for Research in Electronics and Applied Physics; Department of Electrical and Computer Engineering, University of Maryland, College Park, MD}

\author{Christopher Munley}
\affiliation[LPS]
{Laboratory for Physical Sciences, College Park, MD 20740 USA}

\author{Karen E. Grutter}
\affiliation[LPS]
{Laboratory for Physical Sciences, College Park, MD 20740 USA}

\author{Thomas E. Murphy}
\affiliation[UMD]
{Institute for Research in Electronics and Applied Physics; Department of Electrical and Computer Engineering, University of Maryland, College Park, MD}
\email{tem@umd.edu}
\phone{+1 301 405 0030}

\title[Transient Probing of MoS$_2$ on Microresonators]
  {Ultrasensitive Polarization-Resolved Probing of Transient Dynamics in MoS$_2$ on Silicon Nitride Microresonators}

\abbreviations{TMD, NIR, FSR, RIE, LPCVD, TE, TM}


\begin{document}

\begin{abstract}
We present an ultrasensitive technique for probing transient optical changes in atomically thin molybdenum disulfide (MoS$_2$) layers integrated onto silicon nitride (Si$_3$N$_4$) ring resonators. The MoS$_2$ is illuminated by a femtosecond laser, while a tunable near-infrared (NIR) continuous-wave laser probes the microresonator resonance.  The NIR light polarization can be adjusted to either transverse electric (TE, parallel to the 2D material) or transverse magnetic (TM, perpendicular), a configuration that is impossible to achieve with conventional normal-incidence pump-probe techniques. By capturing the transmitted signal on a fast oscilloscope, we detect transient optical shifts with unprecedented sensitivity, observing phenomena over time scales ranging from picoseconds to microseconds. Our results reveal both a rapid, carrier-induced nonlinear optical shift in the resonance, and a slower thermo-optic transient. The ability to simultaneously measure these fast and slow dynamics offers new insight into the complex optoelectronic behavior of 2D materials when integrated with microresonators.  This method provides a significant advance over traditional pump-probe approaches, enabling the detection of exceedingly small transient signals and opening new avenues for exploring the optical properties of atomically thin materials. Our findings highlight the potential of this approach for investigating polarization-dependent nonlinear effects, with applications in photonics, sensing, and optoelectronics.
%
\end{abstract}

 \section*{Introduction}
The integration of two-dimensional (2D) materials with photonic integrated circuits is an emerging approach for introducing active optoelectronic functions into passive photonic components\cite{Datta2020}. This approach leverages the unique properties of 2D materials, such as strong light-matter interaction and large exciton binding energies, within the established framework of integrated optics\cite{Ma2020}. Among the diverse family of 2D materials, transition metal dichalcogenides (TMDs) stand out due to their exceptional optical and electronic characteristics, with molybdenum disulfide (MoS$_2$) being a particularly prominent representative\cite{Singh2020}.

MoS$_2$ exhibits a direct bandgap in its monolayer form, enabling efficient photoluminescence and strong light absorption, which are key for numerous applications\cite{yang2020monolayer}. The atomically thin structure of MoS$_2$ also enhances its nonlinear optical responses, making it a compelling candidate for ultrafast photonic devices\cite{Li2019,srivastava2017mos2,Kar2015}. Moreover, its excitonic properties, including high exciton binding energies, allow for robust excitonic effects even at room temperature\cite{Shi2013, Cha2016}.

Understanding the photophysics of MoS$_2$, particularly its ultrafast carrier dynamics, is essential for realizing its full potential in device applications. These dynamics, which occur on femtosecond to picosecond timescales, involve complex processes such as carrier relaxation, recombination, and intraexcitonic transitions\cite{seo2016ultrafast,Schiettecatte2019,Cha2016}. 
Recent advances in transient absorption and pump-probe techniques have enabled detailed investigations of these dynamics, shedding light on carrier behavior under photoexcitation\cite{Wang2015,Valencia-Acuna2020}.

However, studying the transient optical dynamics of TMDs within such systems presents significant challenges. Conventional normal-incidence pump-probe spectroscopy, a widely used technique for probing ultrafast carrier dynamics in 2D materials, has intrinsic limitations when applied to integrated photonic platforms. One major drawback is its inability to resolve polarization-dependent effects, which are crucial for understanding the anisotropic optical behavior of TMDs \cite{li2020polarization,chen2024characterization}. Additionally, these techniques often lack the spatial resolution needed to probe small-scale integrated systems and struggle to differentiate fast carrier dynamics, occurring on picosecond timescales, and slower thermo-optic responses that evolve over nanoseconds to microseconds\cite{Wang2015}.

The dependence of TMD optical properties on polarization adds another layer of complexity. In integrated photonic platforms, light typically propagates in TE and TM polarized guided modes, which interact differently with the TMD layers \cite{Zhao2022}. As a result, conventional methods fail to capture the full dynamic picture, necessitating advanced experimental methodologies capable of addressing these challenges. By employing approaches that combine temporal resolution over fast and slow timescales with polarization-sensitive detection, one can better identify and quantify the thermal and carrier responses of 2D materials in integrated optical platforms.

In this work, we present a novel polarization-resolved, pump-probe technique for investigating the transient optical response of few-layer MoS$_2$ integrated on Si$_3$N$_4$ microring resonators. Using a femtosecond laser as a photoexcitation source and a tunable near-infrared (NIR) probe, we measure the optically-induced transients for both TE and TM polarization modes. This approach enables the simultaneous characterization of ultrafast, carrier-induced nonlinear optical effects, and slower thermo-optic transients, extending over temporal ranges from picoseconds to microseconds\cite{Datta2024,Tsai2020}.

Our results reveal distinct polarization-dependent transient dynamics in MoS$_2$, providing insight into the interplay between carrier generation, relaxation, and thermal diffusion processes.
The TE and TM mode analyses demonstrate that the carrier-induced nonlinear optical shifts (which exhibit polarization-dependence) dominate the short time scales, while the thermo-optic effects (which are by contrast polarization independent) govern the longer time scales. These findings underscore the importance of polarization-resolved studies in understanding the intrinsic optical anisotropy and dynamic behavior of 2D materials\cite{Singh2020, Wang2019}. The insights derived from this study not only deepen our fundamental understanding of 2D material dynamics but also highlight the potential of Si$_3$N$_4$-based hybrid photonic platforms for applications in optical sensing, nonlinear photonics, and ultrafast optoelectronics. This work contributes to the development of advanced photonic devices, leveraging the unique properties of 2D materials in combination with high-$Q$ resonator architectures\cite{Maiti2020,Marin2019}.

\section*{Results and Discussion}
\subsection*{Device Integration and Experimental Setup}
\begin{figure}[ht]
\centering
  {\includegraphics[width = \textwidth]{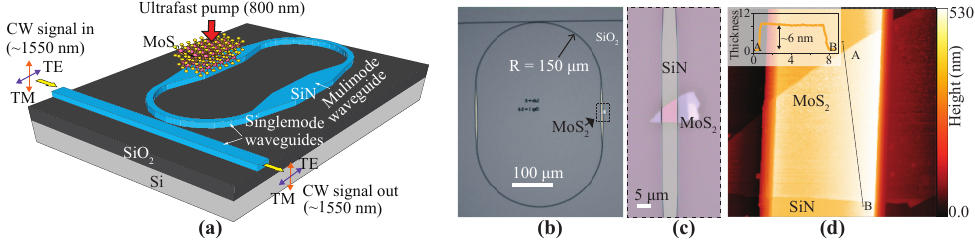}}
\caption{\textbf{(a)} Three-dimensional schematic of the Si\(_3\)N\(_4\) microring resonator integrated with an MoS\(_2\) flake. \textbf{(b)} Optical micrograph of the fabricated microring on a 500~nm Si\(_3\)N\(_4\) layer deposited on a 3~\textmu m SiO\(_2\) undercladding, showing an exfoliated MoS\(_2\) flake (thickness $\sim 6$~nm, corresponding to $\sim9$--$10$ layers) positioned on top of the ring waveguide. \textbf{(c)} Magnified optical micrograph of the MoS\(_2\) region. \textbf{(d)} Atomic force microscopy (AFM) image of the MoS\(_2\) flake.}

\label{scheme}
\end{figure}

Figure \ref{scheme}(a) illustrates a three-dimensional schematic of our integrated racetrack resonator incorporating a thin-film MoS\textsubscript{2} layer. The device is engineered to support both TE and TM fundamental modes, centered around the 1550~nm wavelength of the probe signal. Silicon nitride (Si$_3$N$_4$) is selected as the device layer (500~nm thick) material due to its wide transparency window, extending from the visible to the mid-infrared range, spanning both the probe ($\sim1550$~nm) and pump wavelengths (800~nm). 

To enable integration with a TMD flake covering a larger area than the ultrafast pump laser's spot size, a wide {($5~\upmu$m)}, multimode waveguide section (with mode-size converters at both ends) is incorporated into the resonator. This section facilitates the exfoliation of a uniform TMD flake on the top surface and enhances light-matter interaction at the top relative to the sidewalls (more details on the device integration is given in Section~\ref{sup:dev} of Supporting Information).
 Figures \ref{scheme}(b)-(d) show microscope and AFM images of the fabricated device with a $6$~nm thick (approximately 9-10 layers) MoS\textsubscript{2} flake of $\sim8~\upmu$m length. 
 
\begin{figure}[ht]
\centering
{\includegraphics[width = \textwidth]{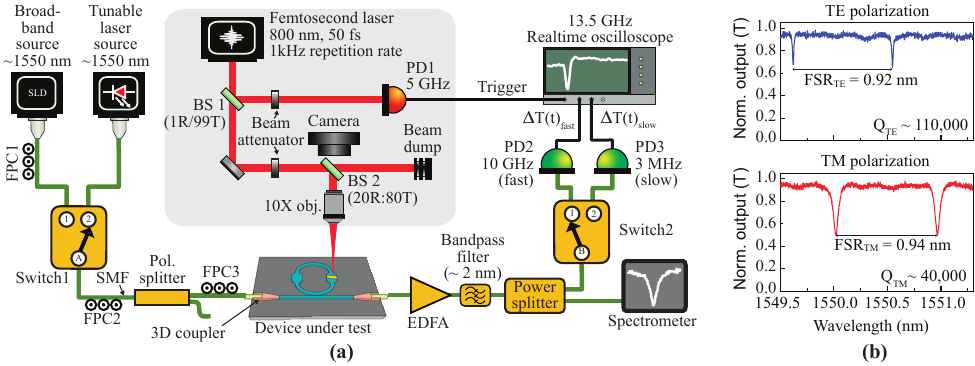}}
\caption{\textbf{(a)} Measurement setup. \textbf{(b)} Normalized transmission spectra for TE (top) and TM (bottom)  polarizations.  FPC - fiber polarization controller, PD - photodetector, BS - beam splitter, SMF - singlemode fiber, SLD - superluminescent diode, EDFA - Erbium doped fiber amplifier.}
\label{setup_exp}
\end{figure}

The experimental setup used to measure the transient response of MoS$_2$ flakes on a microring resonator is shown in Figure \ref{setup_exp}(a). First, the polarization-dependent resonance spectra were measured using a broadband superluminescent diode (SLD) and a spectrometer. 
The normalized transmission spectra ($T(\lambda)$) measured for TE and TM polarizations are shown in Figure~\ref{setup_exp}(b). The two polarization states exhibit distinct free spectral ranges (FSR$_\text{TE}$ and FSR$_\text{TM}$). This difference allowed us to tune the input polarization using a fiber polarization controller and to measure the TE and TM polarization states separately, one at a time.  The loaded quality factor ($Q$) for TE polarization ($Q_\text{TE} \sim 110,000$) is significantly higher than that for TM polarization ($Q_\text{TM} \sim 40,000$).

Subsequently, the input is switched to a continuous-wave (CW) tunable laser source (TLS), which is tuned near the resonance wavelength and serves as the probe signal.
Femtosecond (fs) laser pulses with a central wavelength of 800~nm, a pulse duration of  50~fs, and a repetition rate of 1~kHz were used to photoexcite the MoS$_2$ flakes. While the probe signal remains confined in the integrated waveguide, the pump pulses were focused onto the flake from the out-of-plane direction using a  microscope objective (10$\times$) . A small portion of the pump beam was extracted prior to focusing on the flake and directed to a photodetector (PD1), which provided a trigger signal for oscilloscope synchronization.
The resulting transmitted probe signal was amplified using an erbium-doped fiber amplifier (EDFA) and filtered with a bandpass filter. Two photodetectors with distinct bandwidths were employed to record the photoexcited transients: a 10~GHz photodetector (PD2) captured the fast-varying carrier dynamics, while a 3 MHz photodetector (PD3) monitored the slow-varying thermal effects.  

A 13.6 GHz real-time oscilloscope recorded the transient signals from both detectors, enabling simultaneous observation of the fast and slow transient dynamics. Between each transient measurement, the resonance spectrum was re-measured by switching the optical input from the TLS to the SLD and monitoring the resonance position with a spectrometer. This alternating measurement protocol ensured accurate tracking of resonance wavelength drift caused by ambient temperature fluctuations (see Section~\ref{sup:drift} of Supporting Information).

\begin{figure}[ht]
\centering
  {\includegraphics[width = \textwidth]{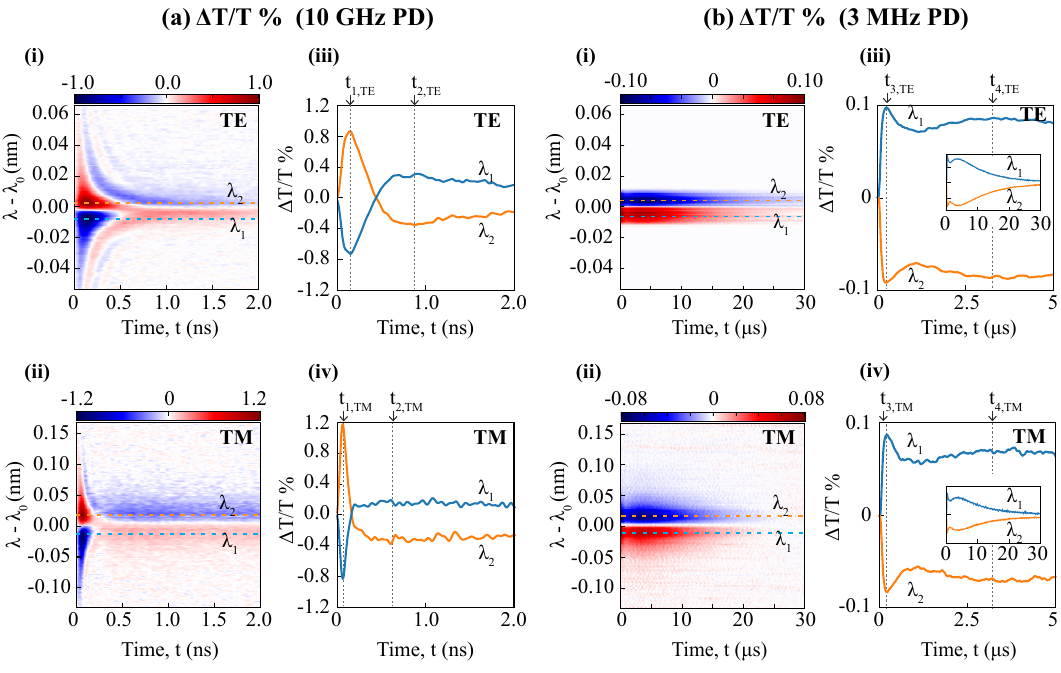}}
\caption{Normalized photoexcited transient changes in transmission ($\Delta T/T$, \%) measured using \textbf{(a)} 10~GHz ($0\leq t \leq2~$ns)  and \textbf{(b)} 3~MHz ($0\leq t \leq30~\upmu$s)  photodetectors. The color plots in (i) and (ii) represent the  transients as a function of wavelength and time for TE and TM polarizations, respectively. (iii) and (iv) represent the corresponding data at two different wavelengths ($\lambda_1<\lambda_r$, $\lambda_2>\lambda_r$) near resonant wavelength ($\lambda_r$). The time axis is shifted to start from 0 by subtracting the time delay between the ultrafast pulse and trigger.}
\label{exp_all}
\end{figure}

The photoexcited transient changes in transmission, $\Delta T / T$ (where $T$ is the off-resonant transmission level, measured at a wavelength sufficiently far from resonance to serve as a baseline), were measured near the resonant wavelengths of $\lambda_r = 1549.6$~nm for TE polarization and $\lambda_r = 1550.05$~nm for TM polarization.
The measurement durations were adjusted for PD2 ($\sim 4~\upmu$s) and PD3 ($\sim 130~\upmu$s) to capture the full transient signal.  The recorded signal amplitudes were also scaled to account for the differing responsivities and amplifier gains of the fast and slow photodetectors.

\subsection*{Transient Measurement Analysis}
The color plots in Figure~\ref{exp_all}(a)(i) and (ii) display the percentage change in transmission ($\Delta T_{\text{PD2}} / T$, \%) for TE and TM polarizations, respectively, over a time interval of 2~ns. To provide further insight, $\Delta T_{\text{PD2}} / T$ (\%) for two distinct wavelengths, $\lambda_1$ ($< \lambda_r$) and $\lambda_2$ ($> \lambda_r$), is plotted in Figure~\ref{exp_all}(a)(iii) for TE polarization and Figure~\ref{exp_all}(a)(iv) for TM polarization. 
The oscillating positive and negative features observed in the fast transients arise partly from the impulse response of the ring resonator. Theory predicts that the response to an intracavity disturbance exhibits oscillations at the optical detuning ($\omega - \omega_0$, where $\omega_0$ is the resonant frequency), with the amplitude of these oscillations decaying on the timescale of the cavity lifetime, as detailed in the Supporting Information.  

For both polarizations, the polarity of the fast transient reverses when the probe laser is tuned from the blue (short wavelength) side to the red (long wavelength) side of the resonance, indicating that the resonance spectrum momentarily blue-shifts toward shorter wavelengths. The time $t_1$ ($t_{1,TE}=160$~ps and $t_{1,TM}=63$~ps) represent the delay at which the peak response was observed.  These times differ not because of an inherently faster material response for TM compared to TE, but rather because the different quality factors of the TE and TM resonators, which result in different cavity decay times.

The initial blue shift in the resonant frequency is consistent with a transient decrease in refractive index due to the generation and subsequent recombination of free carriers, i.e., the free-carrier plasma dispersion effect. The subsequent red shift reflects a pump-induced increase in refractive index, which could arise from thermo-optic or other nonlinear optical processes. Conceptually, the red shift accompanies the blue shift, but because thermo-optic effects evolve more slowly, the observed transient appears as a fast blue-shifted response superimposed on a step-like shift of opposite sign.

The magnitude and temporal dynamics of the blue shift (and the heating time associated with the red-shift) cannot be fully resolved with our detection system, as they are limited by both detector bandwidth and the cavity lifetime. Importantly, we have a detailed characterization of the TE and TM responses, including their extinction ratios and different cavity lifetimes. In the subsequent section, we show that the distinct peak times $t_{1,TE}$ and $t_{1,TM}$ for the blue shift are consistent with a dynamical model that incorporates cavity lifetimes and detector speed, and that by accounting for these differences, we can reliably extract the relative strengths of the TE and TM responses and determine which polarization exhibits the larger effect.

The color plots in Figures~\ref{exp_all}(b)(i) and (b)(ii) illustrate the percentage change in transmission ($\Delta T_{\text{PD3}} / T,\%$) for TE and TM polarizations, respectively, over a time interval of 0--30~$\upmu$s, captured using the 3~MHz photodetector (PD3). The polarity of the slow response is opposite to that of the fast response, indicating a red shift in the resonance toward longer wavelengths due to a thermo-optic increase in refractive index. Unlike the fast response, this signal does not exhibit an initial polarity reversal, as the slower detector lacks the temporal resolution to capture the transient blue shift that occurs on the picosecond timescale. However, the increased sensitivity of the slower detector enables the measurement of smaller, longer-lived thermo-optic shifts that arise from thermal diffusion processes in the waveguide.  
The peak transmission changes associated with the slow thermo-optic effect are notably smaller than those observed for the fast transient response. The peak thermo-optic effect occurs at $t = t_3 ~(\approx260~\text{ns})$, and this timing is the same for both TE and TM polarizations ($t_{3,TE}=t_{3,TM}$), because the thermo-optic transient evolves on a much slower timescale than the cavity lifetime and is therefore unaffected by cavity lifetime limitations.
A magnified view of the transients over a shorter time window (0--8~$\upmu$s) for two distinct wavelengths, $\lambda_1$ ($< \lambda_r$) and $\lambda_2$ ($> \lambda_r$), is presented in Figures~\ref{exp_all}(b)(iii) and (b)(iv) for TE and TM polarizations, respectively. The transient signal exhibits two distinct peaks at times $t_3$ and $t_4$, with the latter peak having a relatively smaller amplitude. Thermal finite-element method (FEM) simulations of the MoS$_2$/waveguide region suggest that the initial peak arises from heating at the MoS$_2$/waveguide interface, which subsequently dissipates into the Si$_3$N$_4$ waveguide, giving rise to the smaller, delayed peak at $t_4$.


It is important to distinguish between the fast nanosecond-scale red shift observed in Figure~\ref{exp_all}(a) and the microsecond-scale thermal relaxation captured by the 3~MHz detector in Figure~\ref{exp_all}(b). The slower response corresponds to the dissipation phase of the thermo-optic effect, governed by both cooling and heat diffusion through the waveguide and cladding layers. In contrast, the earlier red shift likely arises from rapid localized heating of the absorbing 2D material, which can occur on nanosecond timescales when energy is deposited by an ultrafast laser into a nanoscale volume \cite{Khurgin2015}.

The TE and TM resonances differ in both linewidth and extinction ratio due to differences in coupling and intrinsic loss, and these differences must be taken into account when interpreting the relative amplitudes of the measured transient signals.
To extract the underlying optically-induced changes in the ring, we used the framework summarized in the Section~\ref{sup:perturbative} of the Supporting Information, which expresses the change in transmission spectrum, $\Delta T(\lambda)$, as a linear superposition of two known spectral response functions: : one due to a small shift in the resonant wavelength (i.e., a change in round-trip phase), and one due to a small change in round-trip loss. These basis functions, derived analytically from coupled-mode theory, describe how the resonance line-shape shifts or broadens (respectively) in response to small perturbations.  

\begin{figure}[ht]
\centering
  {\includegraphics[width = \textwidth]{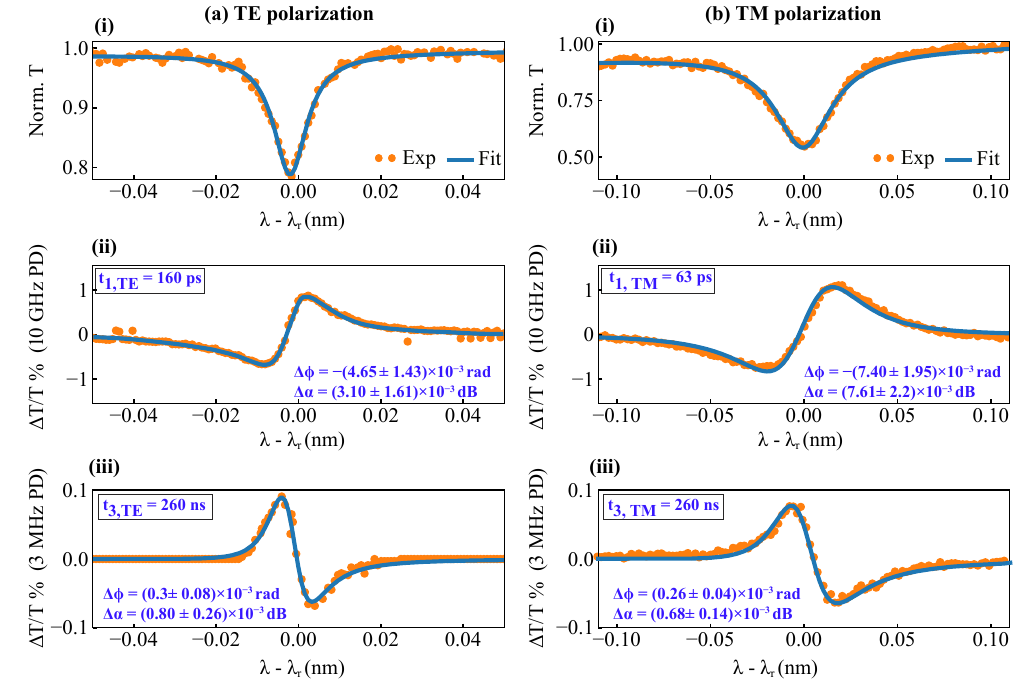}}
\caption{Extracted peak changes in photoexcited transient ($\Delta T/T$, \%) measured across a single resonance for \textbf{(a)} TE and \textbf{(b)} TM polarizations. (i) Unexcited transmission spectra with theoretical fit, (ii) and (iii) represent the ($\Delta T/T$, \%) recorded using fast (10~GHz) and slow (3~MHz) detectors, respectively. Both (ii) and (iii) are fitted with a theoretical perturbative resonator model (see Section~\ref{sup:perturbative} of Supporting Information).}
\label{exp_peakT_fit}
\end{figure}
For a more detailed quantitative analysis of carrier dynamics at the MoS$_2$/waveguide interface, the peak transient changes at $t = t_1$ (blue shift) and $t = t_3$ (red shift) are extracted across the resonance for both polarizations, as shown in Figure~\ref{exp_peakT_fit}(a) and (b). 

Figure~\ref{exp_peakT_fit}(a)(i) and (b)(i) show the normalized transmission spectra for the TE and TM polarizations, respectively, along with their theoretical fits. Figures~\ref{exp_peakT_fit}(a)(ii) and (a)(iii) present the extracted transient peaks at $t_{1,TE}=160~\mathrm{ps}$ and $t_{3,TE}=260~\mathrm{ns}$, respectively. For TM polarization, the corresponding results are shown in Figures~\ref{exp_peakT_fit}(b)(ii) at $t_{1,TM}=63~\mathrm{ps}$ and (b)(iii) at $t_{3,TM}=260~\mathrm{ns}$. While the spectral basis function fitting is strictly valid for quasi-static changes—as at $t_{3}$, where the thermo-optic response can be considered stationary—it is included at the shorter times $t_{1,TE}$ and $t_{1,TM}$ for qualitative comparison, with the understanding that the early-time response is dominated by the system’s dynamical impulse response rather than a purely static resonance shift.

The measured peak transient amplitudes are $\Delta T_{\text{PD2}} / T \approx 1$~\% (fast response) and $\Delta T_{\text{PD3}} / T \approx 0.2$~\% (slow response) for TE polarization, and $\approx 1.2$~\% (fast response) and $\approx 0.1$~\% (slow response) for TM polarization. 
A detailed analysis was performed to extract the effective photoexcited changes in the resonator’s roundtrip phase ($\Delta\phi$) and transmission ($\delta a$) by modeling the transmission spectrum with small perturbations (See Section~\ref{sup:perturbative} of the Supporting Information). As a baseline, the unexcited resonator parameters—power coupling coefficient ($k^2$), round-trip transmission ($a$), and effective index ($n_\text{eff}$)—were first extracted from measurements on the device without the 2D flake. This step establishes the intrinsic waveguide loss ($\alpha$), $k^2$, and $n_\text{eff}$, which are then compared with the corresponding parameters in the presence of the flake. This comparison isolates the flake-induced changes under unexcited conditions, allowing us to more accurately quantify the additional parameter shifts that occur upon photoexcitation.
These results (see Table~\ref{sup:tab1} in the Supporting Information) were subsequently used to fit the photoexcited $T$ and $\Delta T / T$ simultaneously using the \texttt{Bumps}~\cite{kienzle2018bumps} curve fitting tool, assuming that the coupling parameter $r$ remains unchanged (since the coupler is positioned away from the photoexcited region) and accounting for uncertainties in the measured power levels. The corresponding fitted curves for (ii) $\Delta T/T$ at $t = t_1$ and (iii) $\Delta T/T$ at $t = t_3$ are shown in Figure~\ref{exp_peakT_fit}(a) and (b) for TE and TM polarizations, respectively.
The insets show the extracted phase shift $\Delta\phi$ and the change in loss,  
$\Delta \alpha = 10 \cdot \log_{10}(\delta a^2)$ obtained by simultaneously 
fitting the transmission spectra and $\Delta T/T$ data. 
The high-speed response is primarily governed by the refractive index change, with $\Delta\phi_\text{TM}\approx 1.6 \times \Delta\phi_\text{TE}$, rather than by absorption, as indicated by the nearly identical attenuation factors ($\delta a_\text{TE}\approx 0.9996$ and $\delta a_\text{TM}\approx 0.9991$). Although these fits suggest that the TM response is stronger, they do not take into account the temporal dynamics that arise from the different photon lifetimes for TE and TM modes, which we detail in the next section.

\subsection*{Comparison of TE and TM Dynamics}

The TE and TM resonances exhibit different quality factors and extinction ratios, and as a result, the same transient change in the material or waveguide would produce different apparent responses for TE and TM.  However, these differences were accounted for in the fits shown in Fig.~\ref{exp_peakT_fit}.  For the slower dynamics (ns to $\mathrm{\upmu}$s timescales), we find that although the TE and TM transients differ in amplitude, the inferred underlying phase change is largely polarization independent, consistent with expectations for a thermo-optic effect.

For the faster transients, the measured response is determined not only by the material perturbation but also by the photon lifetime in the cavity. A meaningful comparison between TE and TM therefore requires correcting for the differences in their respective temporal responses.  Moreover, if we assume that the observed transient is spatially localized to the 2D material region, then differences in the overlap of the TE and TM modes with this region must also be taken into account.

To better compare the temporal dynamics of the TE and TM resonances, we made a simple model to calculate the time-dependent response of these cavities to a fixed impulse change in the cavity effective index convolved with the ideal Gaussian filter response of the high-speed photodetector (PD2). Details of this model can be found in the Section~\ref{sup:impulse} of Supporting Information. 

\begin{figure}[ht]
\centering
  {\includegraphics[width = \textwidth]{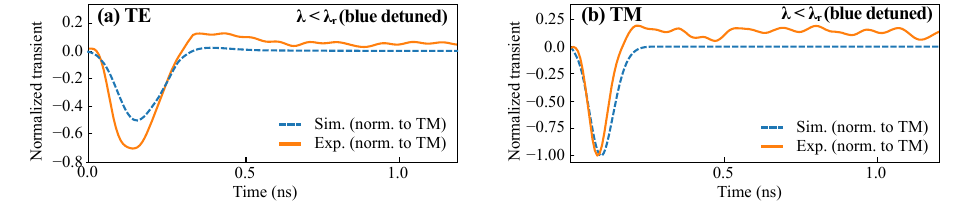}}
\caption{Simulated and experimental transients for (a) TE and (b) TM polarizations, normalized to their respective TM responses. The operating wavelength is detuned near the FWHM ($\lambda < \lambda_r$).}
\label{fig:impulse}
\end{figure}

Figure \ref{fig:impulse} compares simulated and experimental transients at a detuned wavelength near the FWHM ($\lambda < \lambda_r$) for (a) TE and (b) TM polarizations. The experimental data were acquired under identical fluence conditions, and both TE and TM traces were scaled such that the TM transient has a normalized amplitude of 1. In the simulations, the perturbation was modeled as an infinitesimal impulse in the round-trip phase, initially assumed to be the same for TE and TM. The model accounts for differences in quality factor and extinction ratio between the two polarizations, and the simulated response was convolved with a Gaussian temporal function to capture the finite bandwidth of the photoreceiver and oscilloscope. The simulated transients were then normalized in the same way, with the TM response set to 1.  Notably, the simulations correctly predict the approximate times at which the peak transient occurs for both TE and TM cases. 

As shown in Fig.~\ref{fig:impulse}, when the TM simulations are scaled to match the TM experiment, the simulated TE transient underestimates the measured TE response. Conversely, scaling the TE simulation to match experiment would overpredict the TM response. This discrepancy indicates that for the same optical excitation, the transient free-carrier response of the material is stronger for TE polarization, with the in-plane field coupling approximately 1.4$\times$ more effectively to the photoexcited carriers than the out-of-plane TM field. This enhanced TE response is observed after accounting for all differences in cavity lifetime, extinction ratio, and detection-system bandwidth, indicating that it reflects an intrinsic material property. 

\begin{figure}[ht]
\centering
  {\includegraphics[width = \textwidth]{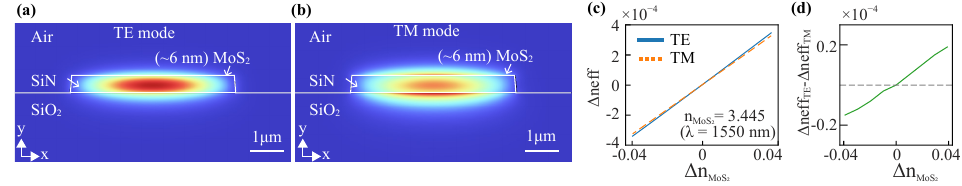}}
\caption{FEM-simulated fundamental (a) TE and (b) TM mode profiles of a Si$_3$N$_4$-on-SiO$_2$ waveguide ($0.5~\upmu\text{m}\times5~\upmu\text{m}$) with a $\sim6$~nm-thick MoS$_2$ layer covering the waveguide. (c) Simulated effective index variation as a function of the MoS$_2$ refractive index for TE ($\Delta\text{neff}_\text{TE}$) and TM ($\Delta\text{neff}_\text{TM}$) polarizations. (d) $\Delta\text{neff}_\text{TE}-\Delta\text{neff}_\text{TM}$. The refractive index of MoS$_2$ at 1550~nm is assumed to be $n_\mathrm{MoS_2}=3.4553$~\cite{song2019layer}.
}
\label{ModeProfile}
\end{figure}

One explanation for the polarization dependence could be a difference in mode overlap with the flake for the two waveguide mode polarizations. To quantify this effect, we calculated the effective mode index using the finite-difference eigenmode (FDE) method, assuming $n_{\text{MoS}_2}=3.4553$ (10 layers) at 1550~nm \cite{polyanskiy2024refractiveindex,song2019layer}. The fundamental TE and TM mode profiles are shown in Figure \ref{ModeProfile}(a) and (b), respectively. By numerically perturbing $n_{\text{MoS}_2}$ in this simulation, we extract $\frac{\partial n_\text{eff}}{\partial n_{\text{MoS}_2}}$ for both TE and TM polarizations (Figure \ref{ModeProfile}(c)). As shown in Figure~\ref{ModeProfile}(d), the difference in these values remains small ($<2\times10^{-5}$), indicating that mode overlap does not contribute significantly to the observed dependence on polarization.

Instead, this suggests that the explanation for the elevated TE response stems from electric field orientation. When the electric field of the waveguide mode is in-plane with the MoS$_2$ flake layers (i.e., the TE mode), it interacts more strongly with the photoexcited carriers than when the electric field is perpendicular to those layers. The same polarization dependence has been observed with dopant-induced phase shift measurements of monolayer WS$_2$~\cite{datta2020low}.

\section*{Conclusions}
We have demonstrated an ultrasensitive, polarization-resolved pump-probe technique for investigating transient optical dynamics in atomically thin MoS$_2$ layers integrated onto silicon nitride microring resonators. This approach represents a significant advancement over conventional normal-incidence pump-probe spectroscopy by enabling simultaneous characterization of both TE and TM polarization modes, a capability that is impossible to achieve with traditional techniques due to the in-plane geometry of integrated photonic platforms.
The exceptional sensitivity of our method, capable of detecting transmission changes as small as 0.1$\%$ corresponding to phase shifts on the order of $10^{-3}$ radians, stems from the high quality factors of the microresonators ($Q > 40,000$) and the enhanced light-matter interaction within the resonant cavity. This sensitivity surpasses that of conventional pump-probe techniques by orders of magnitude, enabling the detection of exceedingly weak transient signals that would otherwise be buried in noise.
Our polarization-resolved measurements reveal distinct anisotropic responses in MoS$_2$, with TE polarization exhibiting approximately 1.4 times stronger carrier-induced phase shifts compared to TM polarization, attributed to the enhanced in-plane electric field interaction with the photoexcited carriers in the  2D material.

The technique successfully separates ultrafast carrier dynamics and cavity damping effects occurring on picosecond-to-nanosecond timescales from the thermo-optic transients extending over microseconds. This temporal separation is crucial for correctly interpreting the physical mechanisms underlying the observed spectral shifts.
The ability to quantitatively extract both phase ($\Delta\phi$) and amplitude ($\delta a$) changes for each polarization provides a direct means to study the fundamental optical anisotropy of transition metal dichalcogenides and their integration with photonic circuits. These findings highlight the potential of silicon nitride-based hybrid photonic platforms for applications in ultrafast optical switching, nonlinear photonics, and high-sensitivity optical sensing, where the polarization-dependent response of 2D materials can be exploited for enhanced device performance.
This work establishes a powerful experimental framework for exploring the rich physics of 2D materials in integrated photonic environments, paving the way for the development of next-generation optoelectronic devices that leverage the unique properties of atomically thin materials in combination with high-$Q$ resonator architectures.

\section*{Methods}

\textbf{Resonator Fabrication and MoS$_2$ Integration:}
We fabricated racetrack ring resonators (150~\textmu m radius) using 500~nm-thick LPCVD Si\textsubscript{3}N\textsubscript{4} on 3~\textmu m-thick thermal SiO\textsubscript{2} (with Si substrate). Photonic structures were defined using electron-beam lithography and inductively coupled plasma reactive ion etching. 
Then, aligned 3D coupler structures with fiber receptacles were created through direct laser writing, employing two-photon polymerization of IP-Dip2 resin~\cite{kudalippalliyalil20243d}. 
Finally, flakes of MoS$_2$ were mechanically exfoliated, transferred to the racetrack ring waveguide, and cleaned using the method described by Le, \emph{et al.}~\cite{le2025assembly}

\begin{suppinfo}

\textcolor{black}{See Supporting Information}

\end{suppinfo}

\bibliography{TMD-Waveguide-Resonator_main}

\newpage
\appendix

\begin{center} {{\bfseries \sffamily \LARGE Supporting Information}}\end{center}

\setcounter{figure}{0}
\makeatletter
\renewcommand{\thefigure}{S\@arabic\c@figure}
\renewcommand{\thetable}{S\arabic{table}}
\setcounter{table}{0} 

\setcounter{equation}{0}
\makeatletter
\renewcommand{\theequation}{S\@arabic\c@equation}

\setcounter{section}{0}
\makeatletter
\renewcommand{\thesection}{S\@arabic\c@section}

\section{Device Design}
\label{sup:dev}
\begin{figure}[!h]
\centering
 {\includegraphics[width = \textwidth]{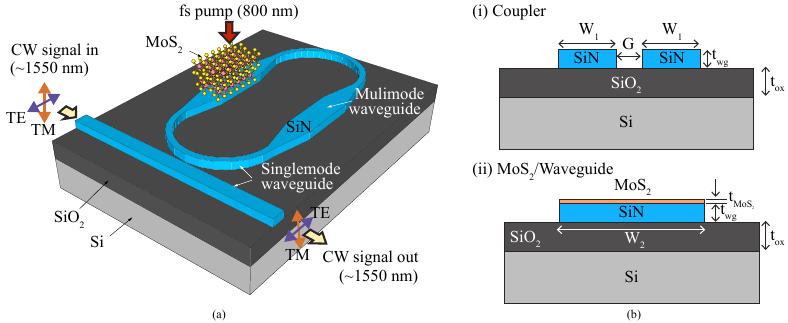}}
\caption{(a) 3D Scheme of the MoS$_2$ integrated SiN racetrack ring resonator. (b) Cross-sections of the (i) coupler and (ii) MoS$_2$/waveguide regions}
\label{S1:device_scheme}
\end{figure}

Figure \ref{S1:device_scheme} shows (a) a three-dimensional schematic of the device and (b) cross-sectional views of the coupler region and the waveguide-integrated MoS$_2$ flake region.
To operate the resonator for both TE and TM polarizations, the width of the coupler waveguides is set to \( W_1 = 1.0 \,\upmu \text{m} \), with a gap of \( G = 0.6 \,\upmu \text{m} \) between them. A bend radius of 150 \(\upmu \text{m} \) is employed to minimize bend losses. The wider multimode waveguides on either side of the ring are designed with a width of 5 \(\upmu \text{m} \), extending over a length of 50 \(\upmu \text{m} \). To suppress higher-order mode excitation, a 45 \(\upmu \text{m}\)-long tapered waveguide is incorporated between the single-mode and multimode sections. The inherent birefringence of the waveguide material induces polarization-dependent optical properties, resulting in distinct group indices \( \left( n_\text{g} = n_\text{eff} - \lambda \frac{d n_\text{eff}}{d \lambda} \right) \), where \( n_\text{eff} \) is the effective index at operating wavelength \( \lambda \), coupling coefficients \( (k^2) \), and round-trip losses \( (\alpha) \) for the TE and TM modes. Consequently, the transmission spectra (see Figure 2(b) of the main text) exhibit unequal free spectral ranges (FSR$_\text{TE}$ = 0.926 nm, FSR$_\text{TM}$ = 0.943 nm) and extinction ratios (ER$_\text{TE}$ $\sim$ 1.7 dB and ER$_\text{TM}$ = 3.2 dB)  for the two polarizations. 


\section{Perturbative Analysis of Ring Resonators}
\label{sup:perturbative}

The transmission through a waveguide that is side-coupled to a traveling wave resonator is
\begin{equation}
T = \frac{a^2 + r^2 - 2ra\cos\phi}{1 + a^2r^2 - 2ra\cos\phi}
\label{eq:mrr_transmission}
\end{equation}
where the dimensionless parameter $r$ characterizes coupling between the waveguide and the resonator: $r^2$ is fractional power that remains in the resonator (or bus waveguide) during a single-pass through the coupler, and $1 - r^2 (= k^2)$ is the fractional power coupled out of the resonator in each round-trip. $\phi$ denotes the phase delay that occurs in one round trip of the resonator, and $a$ denotes the field attenuation factor occurring in one round trip:
\begin{equation}
\phi = \oint \beta(\lambda, z)dz, \quad a = \exp\left(-\oint \alpha(\lambda, z)dz\right)
\end{equation}

When $\phi = 2\pi m$, we have
\begin{equation}
T_{\min} = \left(\frac{a - r}{1 - ar}\right)^2 = 1 - \frac{(1 - a^2)(1 - r^2)}{(1 - ar)^2} \equiv 1 - T_0
\end{equation}
where $0 \leq T_0 \leq 1$ represents the depth of the resonance observed in transmission,
\begin{equation}
T_0 \equiv \frac{(1 - a^2)(1 - r^2)}{(1 - ar)^2}
\end{equation}

For wavelengths sufficiently close to the resonance condition, we can Taylor expand:
\begin{equation}
\phi = 2\pi m + \delta, \quad \cos\phi = 1 - \frac{1}{2}\delta^2 + O(\delta^4)
\end{equation}
from which we get
\begin{equation}
T(\delta) = \frac{(a - r)^2 + ra\delta^2}{(1 - ar)^2 + ra\delta^2} = 1 - \frac{\Gamma^2}{\delta^2 + \Gamma^2}T_0
\end{equation}
where
\begin{equation}
\Gamma^2 \equiv \frac{(1 - ar)^2}{ar}, \quad T_0 \equiv \frac{(1 - a^2)(1 - r^2)}{(1 - ar)^2}
\end{equation}

The phase $\delta$ is related to the (vacuum) wavelength by:
\begin{equation}
\delta = \frac{2\pi}{\Delta\lambda_\text{FSR}}(\lambda - \lambda_m)
\end{equation}
where $\lambda_m$ is the vacuum wavelength of the $m$-th resonance, defined by
\begin{equation}
2\pi m = \oint \beta(\lambda_m, z)dz
\end{equation}
and $\Delta\lambda_\text{FSR} = \lambda^2/(n_gL)$ is the free-spectral range ($n_g$ is the group index)

Note that the resonance is characterized by a Lorentzian spectral dip in transmission of depth $T_0$, with spectral width $\Gamma$. By fitting the spectrum to this functional form, one can obtain information the coupling coefficient $r$ and attenuation factor $a$. However, the two Lorentzian parameters expressed in (7) are symmetric with respect to interchange $a \leftrightarrow r$, which means that $a$ and $r$ cannot be uniquely determined from the transmission spectrum. In order to determine $a$ from a spectral fit, you must know a priori or infer whether the resonator is undercoupled ($r > a$) or overcoupled ($r < a$).

\subsection{Perturbation}

First, suppose that the resonator is perturbed by some small change in the phase in one round trip:
\begin{equation}
\tilde{\phi} = \phi + \Delta\phi
\end{equation}

In this case, the round trip phase is
\begin{equation}
\phi = 2\pi m + \Delta\phi + \delta
\end{equation}
where as before, $\delta$ represents the deviation from the unperturbed resonance condition.

The transmission through the perturbed device is then
\begin{equation}
\tilde{T} \approx 1 - \frac{\Gamma^2}{(\delta + \Delta\phi)^2 + \Gamma^2}T_0
\end{equation}

That is, the induced phase shift causes a spectral shift in the resonance. If the shift $\Delta\phi$ is small in comparison to the linewidth $\Gamma$, we can approximate
\begin{equation}
\Delta T = \tilde{T} - T = \frac{\partial T}{\partial \delta}\Delta\phi = \frac{2\Gamma^2 T_0 \delta}{(\delta^2 + \Gamma^2)^2}\Delta\phi
\end{equation}
\begin{equation}
\Delta T = \frac{2(1 - a^2)(1 - r^2)}{(1 - ar)^2}\frac{\Gamma^3\delta}{(\delta^2 + \Gamma^2)^2}\Delta\phi
\end{equation}
\begin{equation}
\Delta T = f(\delta/\Gamma)\frac{\Delta\phi}{\Gamma}, \quad f(u) = \frac{2(1 - a^2)(1 - r^2)}{(1 - ar)^2}\frac{u}{(1 + u^2)^2}
\label{eqn:phase_perturbation}
\end{equation}

We note that the basis function $f(u)$ is an antisymmetric function of $u$ that is completely determined by the two free parameters $a$ and $r$. 

Now suppose the resonator is instead perturbed by a small change in the amplitude in one round trip:
\begin{equation}
\tilde{a} = a + \Delta a
\end{equation}

This change in the absorption factor can be related to a change in the field attenuation coefficient $\alpha(z)$ that occurs in one round-trip in the resonator:
\begin{align}
\tilde{a} &= \exp\left(-\oint (\alpha(\lambda, z) + \Delta\alpha(\lambda, z))dz\right) \\
&= \exp\left(-\oint \alpha(\lambda, z)dz\right)\exp\left(-\oint \Delta\alpha(\lambda, z)dz\right) \\
&= \exp\left(-\oint \alpha(\lambda, z)dz\right)\left(1 - \oint \Delta\alpha(\lambda, z)dz\right) \\
&= a - a\oint \Delta\alpha(\lambda, z)dz \equiv a + \Delta a
\end{align}

And therefore, the cumulative change in the round-trip absorption is related to $\Delta a$ by:
\begin{equation}
\oint \Delta\alpha(\lambda, z)dz = -\frac{\Delta a}{a}
\end{equation}

The resulting change in transmission can be calculated as
\begin{equation}
\Delta T = \frac{\partial T}{\partial a}\Delta a
\end{equation}

After some algebra, the partial derivative is evaluated to be:
\begin{equation}
\frac{\partial T}{\partial a} = \frac{1 - r^2}{(1 - ar)^3}\left[\frac{(1 + ar)(1 - a^2)}{a}\frac{(\delta/\Gamma)^2}{(1 + (\delta/\Gamma)^2)^2} + 2(a - r)\frac{1}{1 + (\delta/\Gamma)^2}\right]
\end{equation}

When there is a small change $\Delta a$ in the round-trip amplitude, the resulting transient spectral change, $\Delta T$ is given by
\begin{equation}
\Delta T = g(\delta/\Gamma)\Delta a,
\end{equation}
\begin{equation}
g(u) = \frac{1 - r^2}{(1 - ar)^3}\left[\frac{(1 + ar)(1 - a^2)}{a}\frac{u^2}{(1 + u^2)^2} + 2(a - r)\frac{1}{1 + u^2}\right]
\label{eqn:loss_perturbation}
\end{equation}

The basis function $g(u)$ is a symmetric function of $u$ that is completely determined by the two free parameters $a$ and $r$.

The change in propagation loss, expressed in decibels, can be written as
\begin{equation}
\Delta \alpha ~\text{(dB)} = \left |10 \cdot \log_{10}\left(\frac{\tilde{a}}{a}\right)\right|
= \left |10 \cdot \log_{10}(\delta a)\right |,
\label{eqn:lossdB}
\end{equation}
where $\delta a = \tilde{a}/a = 1 + \Delta a / a$. In practice, $\delta a < 1$ because photoexcitation introduces additional absorption due to free carriers.


\section{Extraction of resonator parameters}
\label{sup:resparams}
\subsection{Un-excited MoS$_2$ flake}
Both the photoexcited carrier effects (fast) and thermo-optic (slow) transient responses result in changes to the real and imaginary parts of the effective index in the MoS\textsubscript{2}-integrated waveguide region. To extract these parameters, we fit the transmission spectrum of the resonator under three conditions: (1) before MoS\textsubscript{2} integration ($T_\text{0}$), (2) after MoS\textsubscript{2} integration ($T_{\text{MoS}_2}$), and (3) after photoexcitation ($T_\text{exc}$), using the theoretical model of the resonator.

\begin{figure}[!h]
\centering
  {\includegraphics[width = \textwidth]{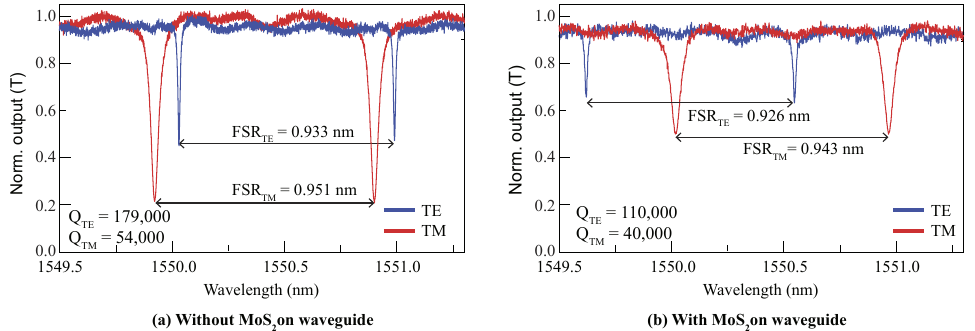}}
\caption{Normalized transmission characteristics of the resonator (a) without and (b) with MoS\textsubscript{2} flake. }
\label{spectra}
\end{figure}

First, the power coupling coefficient ($k^2$) and the round-trip loss ($\alpha$) of the resonator were extracted by fitting the measured transmission spectrum of the resonator without the 2D flakes to the theoretical model (Eq.~\ref{eq:mrr_transmission}).

\begin{table}[ht]
\centering
\caption{Estimated parameters of the ring resonator obtained by fitting the measured resonance spectrum to the theoretical model. $\Delta \alpha_\text{2D}$ and $\Delta n_\text{eff}$ represent the changes in loss and effective index, respectively, calculated over the interaction length $\Delta L \approx 8 \upmu$m of MoS$_2$.}
\begin{tabular}{|c|c|c|c|c|}
\hline
\multirow{2}{*}{\textbf{Parameters}} &\multicolumn{2}{c|}{\textbf{TE}} &\multicolumn{2}{c|}{\textbf{TM}} \\ \cline{2-5}
                                   & \textbf{without MoS\textsubscript{2}}   & \textbf{with MoS\textsubscript{2}}   & \textbf{without MoS\textsubscript{2}}   & \textbf{with MoS\textsubscript{2}}   \\ \hline
FSR (exp.) (nm)       & 0.933      & 0.926      & 0.951       & 0.943        \\ \hline
$Q$ (loaded) & 178,900        & 110,700       & 53,500       & 40,400      \\ \hline
$r$        & 0.9953       & 0.9953        & 0.9284      & 0.9284       \\ \hline
$\alpha$ (dB/cm)                            & 1.56        & 1.56       & 1.98       & 1.98       \\ \hline
$\Delta n_\text{eff}$   & -        &0.0810      & -      & 0.0105        \\ \hline
$\Delta\alpha_\text{2D}$ (dB/cm) & -       & 134        & -       & 156       \\ \hline
\end{tabular}
\label{sup:tab1}
\end{table}

Figure \ref{spectra}(a) shows the transmission spectra for TE and TM polarizations measured before the 2D flake transfer ($T_0$). Figure \ref{spectra} shows the transmission spectra for TE and TM polarizations measured before the 2D flake transfer ($T_0$) and after the 2D flake transfer ($T_\text{MoS2}$). First, $T_0$ was fit to the theoretical model. As initial parameters for the fit, we used the simulated values of the effective indices for single-mode and multimode waveguides. Additionally, the effective coupling length due to bend-induced coupling at the coupler region was calculated for both polarizations. The fit results, summarized in Table~\ref{sup:tab1}, confirm that the TE mode is undercoupled ($r > a$), while the TM mode is overcoupled ($r < a$). This coupling disparity results in a higher quality factor for the TE mode ($Q_\text{TE}$) compared to the TM mode ($Q_\text{TM}$).

The 2D flake has a uniform thickness over the overlapping region $\Delta L = 8 ~\upmu$m, the degree of perturbation to the propagating optical mode at any wavelength depends on the mode overlap with the flake and resulting wavelength-dependent refractive index and absorption characteristics of the material. The material absorption is negligible for range of operating wavelength ($\sim 1550$~nm). 
A change in the effective index ($\Delta n_\text{eff}$) results in a resonance wavelength shift $\Delta \lambda_r \propto \Delta n_\text{eff} \cdot \Delta L$ (a red shift for positive $\Delta n_\text{eff}$ and a blue shift for negative $\Delta n_\text{eff}$). Similarly, a change in the round-trip loss ($\Delta \alpha_\text{2D}$) causes variations in the extinction ratio and the quality factor $Q$ of the resonances. The resonance spectrum after the 2D flake transfer ($T_\text{MoS2}$) is shown in Figure \ref{spectra}(b). Assuming the coupler and waveguide parameters are unchanged, we fit the unexcited transmission spectrum of the 2D-flake integrated resonator, and the estimated values of $\Delta n_\text{eff}$ are $\Delta \alpha$ given in Table~\ref{sup:tab1}.

\subsection{Photoexcited MoS$_2$ flake}
The photoexcited transients are presented in Figure~3 of the manuscript. 
The photo-induced transmission changes are extremely small and cannot be clearly distinguished 
when plotted alongside the unexcited transmission spectra. To analyze these changes, we extracted the peak transmission variations at $t = t_1$ (measured using the 10~GHz photodetector) and $t = t_3$ (measured using the 3~MHz photodetector) across the resonance for both TE and TM polarizations (Figure~4 of the manuscript). 
Using the previously extracted unexcited coupling and loss parameters of the ring, we simultaneously fit the transmission spectra and transient responses 
with the perturbative resonator models (see Figure 4 in the manuscript) in Eq.~\ref{eqn:phase_perturbation} and Eq.~\ref{eqn:loss_perturbation}, using the \texttt{Bumps} curve-fitting tool\cite{ref:kienzle2018bumps1}.

\section{Theoretical Impulse Response Model of the Resonator}
\label{sup:impulse}
The small-signal dynamics of ring resonators are well established \cite{ref:sacher2008dynamics1,ref:pile2014small1}; here, we derive an analytical expression for the impulse response, describing the transmitted power change due to an instantaneous perturbation of the resonant frequency.
\subsection{Coupling of Modes in Space}

Consider a lossless directional coupler that couples two input waves with amplitudes $A$ and $D$ to two output waves with amplitudes $B$ and $C$. The output $C$ is then fed back into the second input $D$ after experiencing a field attenuation factor $a$ and round-trip phase shift $\phi$:

\begin{equation}
    \begin{aligned}
        B &= rA + itD, \\
        C &= itA + rD,
    \end{aligned}
    \label{eq:impulse1}
\end{equation}
where $D = Cae^{i\phi}$ and $r^2 + k^2 = 1$.  

This can be solved to obtain:
\begin{equation}
    B = A (r-\frac{k^2 ae^{i\phi}}{1 - rae^{i\phi}})
      = A (r-\frac{k^2 a}{e^{-i\phi} - ra})
\end{equation}

This expression describes the familiar transmission resonance spectrum, which exhibits periodic dips in transmission that occurs when $\phi = 2\pi m$. Full extinction ($B=0$) occurs at critical coupling when $r=a$.

For wavelengths that are sufficiently close to the resonance condition, we may write:
\begin{equation}
    \phi = 2\pi m + \delta, \quad e^{-i\phi} \approx 1- i\delta,
\end{equation}
which yields:
\begin{equation}
    B = A (r- \frac{k^2 a}{(1 - ra) - i \delta})
    \label{eq:impulse4}
\end{equation}

which describes the spectrum of a single Lorentzian resonance dip in transmission.

\subsection{Coupling of Modes in Time}

Let $V$ represent the optical field in the resonator integrated over one round-trip, normalized such that $|V|^2$ is the total energy in the cavity, while input/output fields $A$ and $B$ correspond to powers $|A|^2$ and $|B|^2$ in the waveguide.

The cavity field is governed by:
\begin{equation}
    \frac{dV}{dt} = -\left(\frac{1}{\tau} - i  \Omega \right)V + i \mu A, ~~~~~B = rA+i\mu V
\end{equation}
where $\Omega = \omega - \omega_m$  is the deviation between the frequency of the input field $\omega$  and the resonant frequency $\omega_m$ of
the ring. $\tau$ is the cavity lifetime (i.e, in the absence of any input field A, the energy
in the cavity will decay with a time constant of $\tau/2$). $\mu$ describes the coupling rate between the resonator and
the side-coupled bus waveguide, which is related to the spatial coupling coefficient $k$ by
$\mu^2 = k^2/T_d$ , where $T_d = L/v_g$ is the round trip group delay of the cavity.

In steady state:
\begin{equation}
    0 = -\left(\frac{1}{\tau} - i \Omega \right)V + i \mu A, ~~~~ B = rA+i\mu A
    \label{6}
\end{equation}
which has the solution:
\begin{equation}
    B = A \left[ r - \frac{\mu^2}{\frac{1}{\tau} - i \Omega} \right].
    \label{7}
\end{equation}

which matches \ref{eq:impulse4}, provided we associate:
\begin{equation}
    \Omega = \delta/T_d, ~~ \mu^2 = k^2/T_d, ~~ 1/\tau = \gamma/T_d
\end{equation}
where $\gamma=(1-ar)$ is related to the spectral width of the resonance.

Now suppose that the resonant frequency $\omega_m$is perturbed from its steady state condition by some small amount $x(t)$:
\begin{equation}
\tilde{\omega_m}(t) = \omega_m + x(t), ~~~ \tilde{\Omega} = \Omega- x(t)
\label{9}
\end{equation}
This will in turn produce a small transient change in $V $and $B$:
\begin{equation}
\tilde{V}(t) = V+v(t), ~~~ \tilde{B} = B+ b(t)
\label{10}
\end{equation}
We note that $A$ is the constant input field amplitude, which is unaffected by this disturbance.Returning to the differential equation for $\tilde{V}$,
\begin{equation}
    \frac{d\tilde{V}}{dt} = -\left( \frac{1}{\tau} - i\tilde{\Omega} \right)\tilde{V} + i\mu A, \qquad 
    \tilde{B} = rA + i\mu \tilde{V}
    \label{11}
\end{equation}

Substituting (\ref{9}) -- (\ref{10}) into (\ref{11}), and subtracting the DC steady-state solution (\ref{6}), 
and ignoring the second-order term proportional to $v(t)x(t)$, we obtain the following 
equation for the small-signal perturbation $v(t)$:

\begin{equation}
    \frac{dv}{dt} + \left( \frac{1}{\tau} - i\Omega \right) v 
    = -iV x(t)
    \tag{12}
\end{equation}

\subsection{Impulse Response}
If $ x(t) = \delta(t)$, the impulse response $v(t)$ is found to be:
\begin{equation}
    v(t) = -iVe^{-t/\tau}e^{i\Omega t}u(t)
\end{equation}
where $u(t)$ is the step function.
The impulse response in the output field is then 
\begin{equation}
     b(t) = i \mu v(t) = \mu V e^{-t/\tau} e^{i \Omega t} u(t)
\end{equation}
but what one observes is the output power, which is given by,
\begin{equation}
     P(t) = |B+b(t)|^2 = |B|^2+2\textbf{Re}\{B^*b(t)\} = P+p(t)
\end{equation}
Therefore the small-signal impulse response in the output power is
\begin{equation}
    p(t) = 2\textbf{Re}\{B^*b(t)\}
\end{equation}

Substituting the solutions above gives
\begin{equation}
    p(t) = 2 |V|^2 \operatorname{Im} \left\{ \left[ r \left( \frac{1}{\tau} - i \Omega \right) - \mu^2 \right] e^{-i \Omega t} \right\} e^{-t/\tau} u(t)
\end{equation}

\begin{equation}
    |V|^2 = \frac{\mu^2}{\dfrac{1}{\tau^2} + \Omega^2} |A|^2
\end{equation}

\begin{equation}
    p(t) = - |A|^2 \, \frac{2 r \mu^2}{\dfrac{1}{\tau^2} + \Omega^2} 
    \left[ \left( \frac{1}{\tau} - \frac{\mu^2}{r} \right) \sin(\Omega t) 
    + \Omega \cos(\Omega t) \right] 
    e^{-t/\tau} u(t)
\end{equation}

In an experiment, it is conventional to normalize the transient response $p(t)$ to the off-resonant transmitted power $P_0$. One can verify from (\ref{7}) that in the limit of large $\Omega$, $P_0 = |B|^2 = |rA|^2$, and so the normalized transient impulse response is

\begin{equation}
\frac{p(t)}{P_0} = -\frac{\frac{2\mu^2}{r}}{\frac{1}{\tau^2} + \Omega^2} \left[ \left( \frac{1}{\tau} - \frac{\mu^2}{r} \right) \sin \Omega t + \Omega \cos \Omega t \right] e^{-t/\tau} u(t) \label{eq:impulse_response}
\end{equation}

where $P_0$ represents transmitted power when the input is detuned far from resonance.

In the experiment, the fixed impulse changes in the cavity response are detected using a high-speed photodetector. 
To compare the experimental results with the theoretical model, the impulse response in Eq.\ref{eq:impulse_response}) is convoluted with a Gaussian filter response that mimics the characteristics of the photodetector.  
The photodetector output can be expressed as:
\begin{equation}
    y(t) = p(t) \circledast g(t),
\end{equation}
where \(g(t)\) is the photodetector Gaussian response given by:
\begin{equation}
    g(t) = \frac{1}{\sqrt{2\pi}\,\sigma} \exp\left[-\frac{(t - t_0)^2}{2\sigma^2}\right],
\end{equation}
where \(\sigma\) is the standard deviation related to the photodetector bandwidth \(B_{\mathrm{PD}}\) by
\begin{equation}
    \sigma = \frac{0.35}{B_{\mathrm{PD}}}.
\end{equation}
Here, \(t_0\) represents the temporal offset (e.g., alignment of the detector response peak), and \(B_{\mathrm{PD}}\) is the \(-3\,\mathrm{dB}\) bandwidth of the photodetector. Thus, the filtered signal \(y(t)\) incorporates both the cavity dynamics and the finite bandwidth of the detection system.
\begin{figure}[!h]
\centering
  {\includegraphics[width = \textwidth]{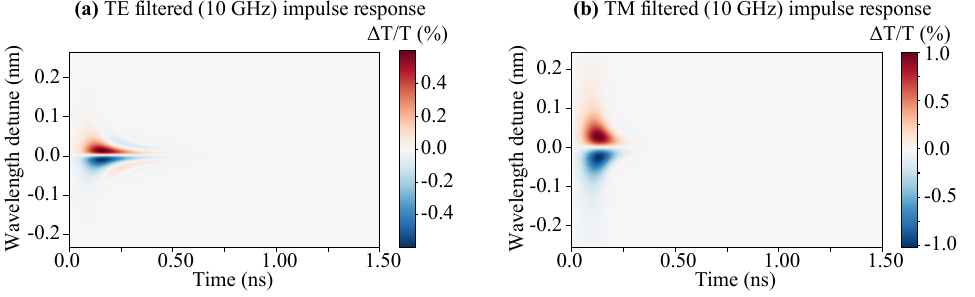}}
\caption{Normalized ($\Delta T/T (\%)$) filtered impulse response for TE and TM polarizations, using a 10~GHz Gaussian filter. Resonator parameters are taken from experimental values listed in Table \ref{sup:tab1}.}
\label{fig:impulse_sweep}
\end{figure}

Figure \ref{fig:impulse_sweep} (a) and (b) show the normalized ($\Delta T/T (\%)$) filtered impulse response for TE and TM polarizations, using a 10 GHz Gaussian filter. Resonator parameters are taken from experimental values listed in Table \ref{sup:tab1}. 
Note that the transient signal exhibits damped oscillatory behavior, corresponding to the sign reversal at $t=t_2$ observed in the experiment. This behavior is characteristic of the impulse response, with the oscillation frequency and damping rate determined by the detuning and cavity lifetime.

\section{Compensating for Drift in the Resonance}
\label{sup:drift}

The high quality factor of the racetrack resonator greatly enhances its sensitivity to the transient response of the photoexcited MoS$_2$ flakes. However, this increased sensitivity also makes the resonator more susceptible to ambient temperature fluctuations in the laboratory, resulting in resonance drift that must be carefully accounted for when interpreting the transient signals. We observed that the resonance drift caused by room temperature variations occurred on a timescale of approximately 15 minutes, whereas each transient measurement was completed in less than 5 seconds. To mitigate this drift, a broadband superluminescent diode source, two optical switches, and a spectrometer were integrated into the experimental setup (see Figure 2), enabling the resonant spectrum of the device to be recorded immediately after each transient measurement. This approach ensured effective drift compensation by continuously monitoring the resonance on a timescale much shorter than that of the ambient temperature-induced drift, allowing these effects to be accurately incorporated into the transient signal analysis.


\end{document}